\documentstyle[epsf,12pt]{article}
\def\doublespace{\lineskip      .25 ex
\baselineskip 5.0 ex
\lineskiplimit 0 ex
\parskip 1.0 ex plus.50 ex minus .25 ex}%
\begin{document}
\doublespace

\centerline{\bf A source of a quasi--spherical space--time:}
\vspace*{0.035truein}
\centerline{\bf The case for the M--Q solution}
\vspace*{0.37truein}
\centerline{{L. Herrera
\footnote{Escuela de F\'{\i}sica, Facultad de Ciencias, Universidad Central de
Venezuela, Caracas, Venezuela. 
Postal address:
Apartado Postal 80793, Caracas 1080A, Venezuela. e--mail:
laherrera@telcel.net.ve}
, W. Barreto}
\footnote{Centro de F\'\i sica Fundamental,
 Departamento de F\'\i sica,
 Facultad de Ciencias, Universidad de los Andes,  M\'erida, Venezuela.
e--mail: wbarreto@ula.ve}
and J.L. Hern\'andez Pastora
\footnote{Departamento de Matem\'atica Aplicada.  
E.T.S.I.I de B\'ejar.\\ Universidad de Salamanca. 
Avda. Fernando Ballesteros, s/n
37700 B\'ejar. Salamanca. Espa\~na. e--mail: jlhp@aida.usal.es}}
\baselineskip=12pt
\vspace*{0.21truein}
\date{\today}

\begin{abstract}
\doublespace
We present a physically reasonable source for an static,
axially--symmetric solution to the Einstein equations. Arguments are
provided, supporting our belief that the exterior space--time produced by
such source, describing a quadrupole correction to the
Schwarzschild metric, is particularly  suitable (among known solutions of the Weyl family)
for discussing the properties of quasi--spherical gravitational fields.

\vspace*{0.21truein}
Key words: Axially--symmetric solutions.
\end{abstract}

\newpage
\doublespace
\section{Introduction}
As is well known, Weyl solutions \cite{2} represent the family of all
static and axially--symmetric exterior solutions to the Einstein equations.
Since there are as many
different Weyl solutions as there are different harmonic functions (see
next section), then the  obvious  question arises: what is the exact vacuum
solution to the Einstein equations
corresponding to a given non--spherical, static axially symmetric source
(an ellipsoid, say)?.

If the field is not particularly intense ($r\gg 2M$)
and the deviation from spherical symmetry is
sligth, then there is not problem in
representing the corresponding field (both inside and outside the source)
as a suitable perturbation of the spherically symmetric exact solution.
However, as the object becomes more
and more compact, such perturbative scheme will eventually fail.

Indeed, as it is well known \cite{1}, the
only static and asymptotically--flat vacuum space--time possessing a
regular
horizon is the Schwarzschild solution. For all the others Weyl exterior
solutions \cite{2}, the physical components of the Riemann tensor
exhibit
singularities at $r=2M$.
 Therefore, it is intuitively
clear that as $r$ approaches $2M$ and
gravitational field becomes stronger, the properties of sources of
Weyl
space--time should start drastically to differ from the properties of spherical
sources \cite{Bel}. It is important to keep in mind that this sharp
difference in
the
behaviour of both types of sources (for very high gravitational fields)
is independent on how small, multipole moments (higher than
monopole)
of the Weyl source, are. This is so because, as $r$ approaches 
$2M$, any finite perturbation of the Schwarzschild space--time becomes
fundamentally different from any Weyl solution, even when the latter is
characterized by parameters whose values are arbitrarily close to those
corresponding to the spherical symmetry. This point has been stressed
before \cite{4}, but usually it has been overlooked.

Furthermore in a recent work \cite{Herrera} it was shown that for a
non--spherical source (even in the case of slight deviations from spherical
symmetry), the speed
of entering the collapse regime decreases substantially, as compared with
the exactly spherically symmetric case. In the same order of ideas it has
been shown \cite{Herreramass}
that small departures from sphericity, produce significant decreasing
(increasing) in the values of active gravitational mass of collapsing
(expanding) spheres, with respect to its
value in  equilibrium, enhancing thereby the stability of the system. Also,
the sensitivity of the trajectories of test particles in the
$\gamma$ spacetime, to small changes of $\gamma$, for orbits close to $2M$,
has been brought out \cite{HS}.
It is important to stress that all these effects take place for strong
gravitational fields, but for $r > 2M$.

These works \cite{Herrera}, \cite{Herreramass}, \cite{HS} were done using
as Weyl solution, the so--called
gamma
metric ($\gamma $--metric) \cite{zipo}. This metric, which is
also
known as Zipoy-Vorhees metric, belongs to the family of Weyl's
solutions, and is continuously linked to the Schwarzschild space--time
through one of its parameters. The motivation for this choice was
that the exterior $\gamma $--metric corresponds to a solution of the
Laplace equation (in cylindrical coordinates) with the same singularity
structure as the Schwarzschild solution (a line segment \cite{zipo}). In
this
sense the $\gamma $--metric appears as the ``natural'' generalization of
Schwarzschild space-time to the axisymmetric case.

However, the existence of so many
different (physically distinguishable \cite{luisyo}) Weyl solutions gives
rise to the question: which among Weyl
solutions is better entitled to describe small deviations from spherical
symmetry?.

Although it should be obvious that such a question does not have a unique
answer (there is an infinite number of ways of being non--spherical, so to
speak), we shall invoke a very
simple criterion, emerging from Newtonian gravity, in order to choose our
solution.

Indeed, in order to answer  the question above, let us recall \cite{yo}
that most known Weyl solutions,  present a drawback  when describing
quasi--spherical space--times. It
consists of the fact that its multipole
structure is such that multipole moments, higher than quadrupole, are of
the same order as the quadrupole moment. Instead, as it is intuitively
clear, the
relevance of such multipole moments
should decrease as we move from lower to higher moments, the quadrupole
moment being the most relevant for a small departure from sphericity. Thus
for example in Newtonian gravity,
multipole moments of an ellipsoid of rotation, with homogeneous density,
and axes (a,a,b) read:
\[ \left\{
\begin{array}{lll}
D_{2n} &= &(-2)^n 3 M a^{2n}\epsilon^n(1-\epsilon/2)^n/[(2n+1)(2n+3)],
\,\,\, \epsilon\equiv (a-b)/a,\nonumber\\
D_{2n+1} &= & 0\nonumber
\end{array}
\right.,\]
because of the factor $\epsilon^n$, this equation clearly exhibits the
progressive decreasing of the relevance of multipole moments as $n$
increases.

Thus, in order to describe small departures from sphericity, by means of
exact solution to the Einstein equations, we  would require an exact solution
whose multipole structure shares the
property mentioned above. Fortunately enough, such solution exists \cite{yo}.
Indeed, there is one (exact) solution  of the Weyl family, which may be
interpreted as a quadrupole correction to the Schwarzschild space--time
(see below).

It is for this exterior metric that we are going to construct a source. The
motivation for this  is twofold: On the one hand, it is always interesting
to propose bounded and
physically reasonable sources of gravitational fields, which may serve as
models of compact object. On the other hand, spherical symmetry is a common
assumption in the
study of compact self--gravitating objects (white dwarfs, neutron stars,
black holes). Therefore it is pertinent to ask, how do small deviations
from this assumption, related to any
kind of perturbation  (e.g. fluctuations of the stellar matter, external
perturbations, etc.), affect the properties of the system?. However, as
mentioned before, for sufficiently
strong fields, in order to answer to this question it is necessary to deal
with non--spherically symmetric exact solution to the Einstein equations.

For constructing the source we shall follow a
prescription given by Hern\'an\--dez \cite{8} allowing to
obtain interior solutions of Weyl space--time, from known spherically
symmetric interior solutions. Our interior solution will be obtained from
the interior Schwarzschild solution (homogeneous density).

The paper is organized as follows. In the next section we review Weyl
solutions and the concept of relativistic multipoles. In Section 3 we
describe the M--Q solution \cite{yo} and give
its properties. The Hern\'andez method is applied in Section 4 to obtain an
interior to the M--Q solution. Finally, results are discussed in the last
Section.

\section{The Weyl metrics}

Static axysymmetric solutions to the Einstein equations are given by the Weyl
metric \cite{2}
\begin{equation}
ds^2 = -e^{2 \Psi} dt^2 + e^{-2 \Psi} [e^{2 \Gamma}(d \rho^2 +dz^2)+\rho^2
d \phi^2 ],
\label{elin}
\end{equation}
where metric functions have to satisfy
\begin{equation}
\Psi_{, \rho \rho}+\rho^{-1} \Psi_{, \rho}+\Psi_{, zz} = 0
\label{meq1}
\end{equation}
and
\begin{equation}
\Gamma_{,\rho}= \rho (\Psi_{, \rho}^2-\Psi_{, z}^2) \quad; \quad \Gamma_{,
z}= 2 \rho \Psi_{,\rho} \Psi_{, z}.
\label{meq2}
\end{equation}

Observe that  (\ref{meq1}) is just the Laplace equation for $\Psi$ (in the
Euclidean space), and furthermore it represents the integrability condition
for (\ref{meq2}), implying that
for any ``Newtonian'' potential we have a specific Weyl metric, a well
known result.

The general solution of the Laplace equation (\ref{meq1}) for the function
$\Psi$, presenting an asymptotically flat behaviour, results to be
\begin{equation}
\Psi = \sum_{n=0}^{\infty} \frac{a_n}{r^{n+1}} P_n(\cos \theta),
\label{psi}
\end{equation}
where $r=(\rho^2+z^2)^{1/2}$, $\cos \theta= z/r$ are Weyl spherical
coordinates and $P_n(\cos \theta)$ are Legendre Polynomyals. The
coefficients $a_n$ are arbitrary real constants
 which have been named in the literature ``Weyl moments'', although they
cannot be identified as relativistic multipole moments in spite of the
formal similarity between expression
(\ref{psi}) and the Newtonian potential.

Another interesting way of writting the solution (\ref{psi}) was obtained
by Erez--Rosen \cite{erroz} and Quevedo \cite{quev}, integrating equations
(\ref{meq1}),  (\ref{meq2})
in prolate spheroidal coordinates, which are defined as follows
\begin{eqnarray}
x_{\,\,} & = & \frac {r_{+}+r_{-}}{2 \sigma} \quad , \quad y  =  \frac
{r_{+}-r_{-}}{2 \sigma},\nonumber \\ \nonumber \\
r_{\pm} & \equiv & [\rho^2+(z\pm \sigma)^2]^{1/2}, \nonumber \\ \nonumber \\
x_{\,\,} & \geq & 1 \quad , \quad -1 \leq y \leq 1,
\label{pro}
\end{eqnarray}
where $\sigma$ is an arbitrary constant which will be identified later with
the Schwarzschild's mass. Inverse relation between both families of
coordinates is given by
\begin{eqnarray}
\rho^2 & = & \sigma^2(x^2-1)(1-y^2), \nonumber \\ \nonumber \\
z & = & \sigma \, x \, y.
\label{invpro}
\end{eqnarray}
The prolate coordinate $x$ represents a radial coordinate, whereas the
other coordinate, $y$ represents the cosine function of the polar angle.

In these prolate spheroidal coordinates  $\Psi$ takes the form
\begin{equation}
\Psi = \sum_{n=0}^{\infty} (-1)^{n+1} q_n Q_n(x) P_n(y),
\label{propsi}
\end{equation}
being $Q_n(y)$ Legendre functions of second kind and $q_n$ a set of
arbitrary constants.
The corresponding expression for the function $\Gamma$, has been obtained
by Quevedo \cite{quev}

A sub--family of Weyl solutions has been obtained by Gutsunaev and Manko
\cite{GM85}, \cite{M89}
starting from the Schwarzchild solution as a ``seed'' solution.

Both sets of coefficients, ${a_n}$ and ${q_n}$, characterize any Weyl
metric \cite{quev}. Nevertheless, as mentioned before, these constants do
not give us physical information about
the metric since they  do not represent the ``real'' multipole moments of
the source. That is not the case for the relativistic multipole moments
firstly defined by Geroch \cite{ger},
and subsequently by Hansen
\cite{han} and Thorne
\cite{th}, which, as it is known, characterize completely and uniquely, at
least in the neighbourhood of infinity, every asymptotycally flat and
stationary vacuum solution
\cite{kun1}, providing at the same time a physical description
of the corresponding solution.

An algorithm to calculate the Geroch multipole moments was developed by
G. Fodor, C. Hoenselaers and Z. Perjes \cite{fhp} (FHP). By applying such
method, the resulting multipole
moments of the solution are expressed in terms of the Weyl moments. Similar
results are obtained from the Thorne's definition, using harmonic
coordinates.  The structure of the
obtained relation between coefficients $a_n$ and these relativistic moments
allows to express the Weyl moments as a  combination of the Geroch
relativistic moments.
\vskip 1cm

\subsection {The Monopole--Quadrupole solution, $M-Q$}

\vskip 2mm

In this section we would like to describe the properties of a solution
(hereafter referred to as the $M-Q$ solution \cite{yo}) which is
particularly suitable for the study of
perturbations of the spherical symmetry. The main argument to support this
statement is based on the fact that previously known Weyl metrics (e.g.
Gutsunaev--Manko \cite{GM85}, Manko
\cite{M89}, gamma metric \cite{zipo}, Curzon \cite{curzon}, etc.) have a
multipolar structure (in the Geroch sense) such that all the moments higher
than the quadrupole, are of the same
order as the quadrupole. In fact for the above mentioned metrics we have (odd
moments are of course vanishing)
\begin{eqnarray}
 M_0^{GM} & = & M_0^{ER} = M, \nonumber\\
 M_2^{GM} & = & M_2^{ER} = \frac{2}{15} q_2 M^3, \nonumber\\
 M_4^{GM} & = & -3 M_4^{ER} = \frac{4}{35} q_2 M^5, \nonumber\\
 M_6^{GM} & = & M_6^{ER}-\frac 27 \frac{817}{33} M^2 M_4^{ER} =
\frac{2}{15} \frac{4}{231} q_2 M^7 (\frac{194}{7}
+\frac{14}{15} q_2),\nonumber
\label{(14)}
\end{eqnarray}
where $q_2$ is the quadrupole parameter in the  Erez-Rosen metric. For the
gamma metric results to be
\begin{eqnarray}
M_0 &=& \gamma M, \nonumber\\
M_2 &=&\gamma \frac{M^3}{3} (1-\gamma^2), \nonumber \\
M_4 &=&\gamma \frac{M^5}{5} (1-\gamma^2) (1-\frac{19}{21} \gamma^2), \nonumber\\
M_6 &=&\gamma \frac{M^7}{7} (1-\gamma^2)
(1+\frac{389}{495}\gamma^4-\frac{292}{165} \gamma^2), \nonumber
\label{MMgamma}
\end{eqnarray}
and finally, Curzon metric is the worst case of the mentioned metrics since
it posseses a unique parameter which represents the mass, and all higher
moments are proportional to increasing powers of that parameter, i.e.,
\begin{eqnarray}
M_0 &=& -a_0, \nonumber\\
M_2 &=& \frac{1}{3}{a_0^3},\nonumber\\
M_4 &=& -\frac{19}{105} a_0^5,\nonumber\\
M_6 &=& \frac{389}{3465} a_0^7.\nonumber
\end{eqnarray}
In \cite{yo} it was shown that it is possible to find a solution of the
Weyl family, by a convenient choice of coefficients
$a_n$, such that the resulting solution possesses only monopole and
quadrupole moments (in the Geroch sense) \cite{ger}. The obtained solution
($M-Q$)
may be written as a finite series of  Gutsunayev--Manko  \cite{GM85}  and
Erez--Rosen \cite{erroz} solutions, as follows:
\begin{equation}
\Psi_{M-Q}=\Psi_{q^0}+q \Psi_{q^1}+q^2 \Psi_{q^2}+\ldots =
\sum_{\alpha=0}^\infty q^\alpha\Psi_{q^\alpha} \quad ,
\label{(15)}
\end{equation}
where the zeroth order corresponde to the
Schwarzschild solution.
\begin{equation}
\Psi_{q^0}=-\sum_{n=0}^{\infty}\frac
{\lambda^{2n+1}}{2n+1}P_{2n}(\cos\theta) \quad ,
\label{(16)}
\end{equation}
with $\lambda \equiv M/r$ and
it appears that each power in $q$ adds a quadrupole correction to the
spherically symmetric solution. Now, it should be observed that due to the
linearity of Laplace equation,
these corrections give rise to a series of exact solutions. In other
words, the power series of $q$ may be cut at any order, and the partial
summatory, up to that order, gives an exact
solution representing a quadrupolar correction to the  Schwarzschild solution.

The simplest way to interpret physically the exact solutions obtained from
the quadrupolar corrections described above, consists in analyzying the
corresponding multipolar structure.
Thus , it can be shown that cutting solution (\ref{(15)}) at some order
$\alpha$,
one obtains an exact solution with the following properties:
\begin{itemize}
\item Both, the monopole and the quadrupole moments are non--vanishing:

 $M_0\equiv M$, $M_2 \equiv q M^3$.

\item All the remaining moments until  order $2(\alpha+1)$  (included) vanish.

\item All moments above the  $2(\alpha+1)$--pole are of order $q^{\alpha+1}$. Therefore,
the solution represents a quadrupolar correction to the Schwarzschild
solution, which is an exact
solution up to the given order.
\end{itemize}
To illustrate further our point, let us present the explicit solution up to
first order, describing a quadrupolar correction to the monopole. The
correspondig metric functions read
(note a misprint in the equation (13) in \cite{yo2})
\begin{eqnarray}
\Psi_{M-Q}^{(1)} &\equiv & \Psi_{q^0}+q \Psi_{q^1}\nonumber\\
& =& \frac12
\ln\left(\frac{x-1}{x+1}\right) + \frac58 q(3y^2-1)\times \nonumber\\
\nonumber
\end{eqnarray}
\begin{equation}
\times\left[\left(\frac{3x^2-1}4 -\frac1{3y^2-1}\right)
\ln\left(\frac{x-1}{x+1}\right)\nonumber\\
 -\frac{2x}{(x^2-y^2)(3y^2-1)} + \frac{3x}2\right],
\label{(25)}
\end{equation}

\begin{eqnarray}
\Gamma^{(1)}_{M-Q} & \equiv & \Gamma_{q^0}+q \Gamma_{q^1}+q^2 \Gamma_{q^2}
=\frac12\left(1+\frac{225}{24}
q^2\right)\ln\left(\frac{x^2-1}{x^2-y^2}\right) \nonumber\\
&+& \frac{225}{1024}q^2(x^2-1)(1-y^2)(x^2+y^2-9x^2y^2-1)
\ln^2\left(\frac{x-1}{x+1}\right) \nonumber\\
&-& \frac{15}4 q(1-y^2)\left[1-\frac{15}{64}q(x^2+4y^2-9x^2y^2+4)\right]
\nonumber\\
&-& \frac{75}{16}q^2 x^2\frac{1-y^2}{x^2-y^2} - \frac{5}{4}q(x^2+y^2)
\frac{1-y^2}{(x^2-y^2)^2} \nonumber\\
&-& \frac{75}{192}q^2(2x^6-x^4+3x^4y^2-6x^2y^2+4x^2y^4-y^4-y^6)
\frac{1-y^2}{(x^2-y^2)^4} \nonumber \\
&-& \frac{15}8 q x(1-y^2)\left[1- \frac{15}{32}q\left(x^2+7y^2-9x^2y^2+1-\frac83
\frac{x^2+1}{x^2-y^2}\right)\right] \nonumber\\
&\times&\ln\left(\frac{x-1}{x+1}\right).
\label{(26)}
\end{eqnarray}

The first twelve Geroch moments of this solution are
(odd moments vanish because of the reflexion symmetry)
\begin{eqnarray}
M_0 &=& M , M_2 = M^3 q , M_4 = 0 ,
M_6 = -\frac{60}{77} M^7 q^2 ,\nonumber\\
\vspace{0.3cm}
M_8 &=& -\frac{1060}{3003} M^9 q^2 - \frac{40}{143} M^9 q^3 ,
M_{10} = -\frac{19880}{138567} M^{11} q^2 +
\frac{146500}{323323} M^{11} q^3 ,\nonumber\\
\vspace{0.3cm}
M_{12}&=& -\frac{23600}{437437}M^{13}q^2 + \frac{517600}{1062347}M^{13}q^3 +
\frac{4259400}{7436429}M^{13}q^4.
\label{(27)}
\end{eqnarray}
From the expressions above, it is apparent that the parameter
$q\equiv M_2/ M^3$ representing the quadrupole moment, enters into the
multipole moments $M_{2n}$, for $n\ge 2$,
only at order $2$ or higher. Accordingly,  solution
(\ref{(25)})--(\ref{(26)}) for an small value
of $q$, up to order $q$, may be interpreted as the gravitational field
outside a quasi--spherical source.
The spacetime being represented by a quadrupole correction to the monopole
(Schwarzschild) solution. This is in contrast with other previuosly
mentioned solutions of Weyl family, where
all moments higher than the quadrupole are of the same order in $q$ as the
quadrupole, and therefore for small values of $q$ they cannot be
interpreted as a quadrupole perturbation of
 spherical symmetry.

\vskip 1cm
Instead of cylindrical corrdinates $(\rho,z)$, it will be useful for the
next section to work with   Erez-Rosen coordinates
$(r,
\theta)$  given by:
\begin{eqnarray}
z &=&(r-M) \cos \theta, \nonumber\\
\rho &=& (r^2-2 M r)^{1/2} \sin\theta,
\label{(28)}
\end{eqnarray}
and related to prolate coordinates, by
\begin{eqnarray}
x &=& \frac{r}{M}-1, \nonumber\\
y &=& \cos\theta.
\label{(29)}
\end{eqnarray}

The metric functions of the  solution, up to the
first order in $q$, hereafter
refered as M-Q$^{(1)}$ are:
\begin{eqnarray}
\Psi_{M-Q}^{(1)}  &  = & \frac12
\ln\left(1-\frac2R\right) + \frac{5}{32} q(3y^2-1)(3 R^2-6
R+2)\ln\left(1-\frac2R\right) \nonumber\\
\vspace{0.2cm}
&-&\frac58 q \ln\left(1-\frac2R\right)-\frac54 q
\frac{R-1}{(R-1)^2-y^2}+\frac{15}{16}q (3 y^2-1) (R-1) \quad ,
\label{(30.1)}
\end{eqnarray}

\begin{eqnarray}
\Gamma^{(1)}_{M-Q} &  = &\frac12
\ln\left[\frac{(R-1)^2-1}{(R-1)^2-y^2}\right]-\frac{15}{8} q
(1-y^2)(R-1)\ln\left(1-\frac2R\right) \nonumber\\
\vspace{0.2cm}
&-&\frac{15}{4} q (1-y^2)-\frac54 q(1-y^2)
\left[\frac{(R-1)^2+y^2}{((R-1)^2-y^2)^2}\right],
\label{(30.2)}
\end{eqnarray}
with $R\equiv r/M$. An study of the geodesics in this spacetime, has been
recently presented \cite{geomq}

In  the next section we shall construct a source for M-Q$^{(1)}$ solution.
\section{An interior M-Q$^{(1)}$ metric}
The Hern\'andez method \cite{8} is based on a  heuristic procedure which
allows, starting with some spherically symmetric ``seed''  source, to
obtain interior solutions  describing
axialsymmetric static sources, which match smoothly on the boundary
surface, to a given metric of the Weyl family (see \cite{8} for details).
Some applications of this method may be
found in \cite{7}.

 For our exterior spacetime we shall
choose the $M-Q^{(1)}$ solution described above (up to first order in $q$)
and our ``seed'' interior fluid
will be the incompressible Schwarzschild interior solution.

Thus, our exterior metric in Erez--Rosen coordinates read:
\begin{eqnarray}
g_{rr} &  = & e^{2 \Gamma-2 \Psi} \left(1+\frac{\lambda^2\sin^2\theta}{1-2
\lambda}\right)=   e^{2q(\Gamma_{q^1}-
\Psi_{q^1})}/(1-2M/r),\nonumber\\
\vspace{0.2cm}
g_{\theta \theta} &  = & e^{2 \Gamma-2 \Psi} r^2 (1-2 \lambda +\lambda^2
\sin^2\theta) = r^2 e^{2q(\Gamma_{q^1}- \Psi_{q^1})}, \nonumber\\
\vspace{0.2cm}
g_{\phi\phi} &  = & e^{-2 \Psi} r^2 \sin^2 \theta(1-2 \lambda) = r^2 \sin^2
\theta  \ e^{-2q\Psi_{q^1}},\nonumber\\
\vspace{0.2cm}
g_{tt} &  = & - e^{2 \Psi} = -(1-2M/r)\  e^{2q\Psi_{q^1}},
\label{(31)}
\end{eqnarray}
where  the metric functions  $\Gamma_{q^1}$ and
$\Psi_{q^1}$
are given by
\begin{eqnarray}
\Psi_{q^1} &  = & \frac{5}{32} (3y^2-1)(3 R^2-6
R+2)\ln\left(1-\frac2R\right) \nonumber\\
\vspace{0.2cm}
&-&\frac58 \ln\left(1-\frac2R\right)-\frac54
\frac{R-1}{(R-1)^2-y^2}+\frac{15}{16} (3 y^2-1) (R-1) \quad ,
\label{(30.1N)}
\end{eqnarray}
\begin{eqnarray}
\Gamma_{q^1} &  = &-\frac{15}{8}
(1-y^2)(R-1)\ln\left(1-\frac2R\right) \nonumber\\
\vspace{0.2cm}
&-&\frac{15}{4} (1-y^2)-\frac54 (1-y^2)
\left[\frac{(R-1)^2+y^2}{((R-1)^2-y^2)^2}\right].
\label{(30.2N)}
\end{eqnarray}

Now, Darmois conditions, in these coordinates, imply that metric components
as well as $g_{\theta \theta ,r}$, $g_{t t ,r}$, $g_{\phi \phi ,r}$ are
continuosus across the boundary surface
(but allows a jump in $g_{r r ,r}$).

Thus, in the example given by Hern\'andez \cite{8}, the following substitutions
on the chosen exterior metric were applied, in order to obtain the interior
metric:
\begin{equation}
2M/r \rightarrow r^2/B^2,
\label{42}
\end{equation}
for the $g_{rr}$ metric component and the following one for the other
metric components
\begin{equation}
2M/r \rightarrow 1-\left[\frac32
\left(1-\frac{r_{\Sigma}^2}{B^2}\right)^{1/2}-\frac12
\left(1-\frac{r^2}{B^2}\right)^{1/2}\right]^2,
\label{43}
\end{equation}
where $r=r_{\Sigma}$ is the equation of the boundary surface of the source
and $B^2=3/(8 \pi \rho_{s})$, with $\rho_{s}$ denoting the energy
density of the spherically symmetric
``seed'' solution. Since (\ref{42}) does not have a continuous derivative
at the surface, but (\ref{43}) has, it is clear that Darmois conditions
will be satisfied by the so obtained
metric, on the boundary surface.

 However, in our case these substitutions lead to a metric whose $g_{rr}$
component is singular at the origin. Thus as a first step, we are going to
modify the
Hern\'andez rule, concerning the
$g_{rr}$ component. In order to have a regular metric at the origin, we
shall use (\ref{43}) in those terms of $g_{rr}$ which produce
the singularity at the origin and (\ref{42}) in the remaining terms. For
the others metric components we apply (\ref{43}). In addition, we modify
the spherical factor in $g_{\theta \theta}$ and $g_{\phi \phi}$ by
substituing $r^2$ with $r^2+q(r-
r_{\Sigma})^2$,
in order to ensure regularity at the origin (however, as we shall see below
this is not enough to assure the correct physical behaviour at the centre).
Thus we obtain,
\begin{equation}
g_{tt}= -X(r)^{1+a(r,y)} e^{b(r,y)},
\label{(34)}
\end{equation}
\begin{equation}
g_{rr}=(1-r^2/B^2)^{-(1+c(r,y))}e^{d(r,y)},
\end{equation}
\begin{equation}
g_{\theta \theta} =  (r^2+q(r-
r_{\Sigma})^2) X(r)^{-c(r,y)}e^{d(r,y)},
\end{equation}
\begin{equation}
g_{\phi\phi}= (r^2+q(r-
r_{\Sigma})^2) \sin^2\theta
 X(r)^{-a(r,y)}e^{-b(r,y)},
\end{equation}
with
\begin{equation}
X(r)  \equiv \left[\frac32 \left(1-\frac{r_{\Sigma}^2}{B^2}\right)^{1/2}-\frac12
\left(1-\frac{r^2}{B^2}\right)^{1/2}\right]^2,
\label{(33II)}
\end{equation}
\begin{equation}
a(r,y)={\frac{15}{8} q \frac{-(1+X(r)^2)(1-y^2)+4 X(r)
y^2}{(1-X(r))^2}},
\end{equation}
\begin{equation}
b(x,y)={\frac52 q \frac{1+X(r)}{1-X(r)} \left[ \frac34 (3
y^2-1)-\frac{(1-X(r))^2}{(1+X(r))^2-y^2(1-X(r))^2}\right]},
\end{equation}
\begin{equation}
c(r,y)={\frac{15}{16}
q\left[4(1-y^2)\frac{1+X(r)}{1-X(r)}+(3y^2-1)\frac{(1+X(r))^2}{(1-X(r))^2}-y
^2-1\right]},
\end{equation}
\begin{eqnarray}
d(r,y)&=&-\frac52 q\left[ 3(1-y^2)+\frac34
(3y^2-1)\frac{1+X(r)}{1-X(r)}\right .\nonumber\\
&+&(1-y^2)(1-X(r))^2\frac{(1+X(r))^2+y^2(1-X(r))^2}{\left[(1+X(r))^2-y^2(1-X
(r))^2\right]^2}\nonumber\\
&-&\left .\frac{1-X(r)^2}{(1+X(r))^2-y^2(1-X(r))^2}\right].
\end{eqnarray}
Next, let us introduce the following dimenssionles variables:
$\alpha=2 M/r_{\Sigma}$ and $\beta=r/r_{\Sigma}$, in terms
of which we may write

\begin{equation}
X(\beta)  \equiv \left[\frac32 \left(1-\alpha \right)^{1/2}-\frac12
\left(1-\beta^2 \alpha \right)^{1/2}\right]^2.
\label{(33)}
\end{equation}

We can now calculate the components of the energy--momentum tensor. However
when this is done, negative values of the  energy density close to the
center appear, even in
the weak field limit. To solve this problem and to assure a correct
physical behaviour of all physical variables, we shall consider the
parameter $q$ as a function of $r$, such
that
\begin{equation}
 q=0; \;\;
 \beta \in [0,\beta_o]
\end{equation}
and
\begin{equation}
q=q_o(\beta-\beta_o)^4 (\beta-\beta_s)^4; \;\;
 \beta \in [\beta_o,(\beta_o+\beta_s)/2],
\end{equation}
where $\beta_o$, $\beta_s$ and $q_o$ are constants such  that
$\beta_o+\beta_s =2$
and
the value  of  $q$ at the boundary surface, which is
$q_o[(\beta_o-\beta_s)/2]^8$ coincides with the quadrupole parameter of the
exterior  M-Q$^{(1)}$ solution, i.e.  $q=q_{\Sigma}$,
 where $q_{\Sigma}$ denotes the quadrupole parameter of the exterior
M-Q$^{(1)}$. It should be observed that since both $q$ and its first
derivative  are continuous across
the boundary surface, junction conditions are satisfied, after the
replacement above. The specific form of $\beta$ as well as the values of
different parameters, are indicated in figure 1 and in its legend.

Thus our source consists of a spherical inner core continuously matched to
a non--spherical distribution of matter, producing a   M-Q$^{(1)}$
spacetime at the outside, and satisfying
the continuity of the first and the second fundamental forms at the
boundary surface.

Calculations of different physical variables show their correct physical
behaviour within the
matter distribution.

Expressions are extremely lengthy and therefore we omit them here, however
they are available upon request to W. Barreto.

The non--vanishing components are $T^0_{0}$, $T^1_{1}$, $T^2_{2}$, $T^3_{3}$,
$T^1_{2}$. For $\alpha < 0.7$ all these components are regular
 within the fluid distribution and energy conditions
are satisfied (e.g. energy density is positive and larger than stresses).
In  the weak field limit ($\alpha \ll 1$) we obtain a quadrupole correction to
the incompressible fluid.

\section{Conclusions}
We have seen that the  M-Q$^{(1)}$ solution satisfies the requested
condition to be considered as a  quadrupole correction to  the spherical
symmetry, namely: all relativistic moments
higher than the quadrupole are of higher order in $q$. Accordingly it
represents, among the known members of the Weyl family, a good candidate
to describe small departures from
spherical symmetry.

Next, we have found a source for that space--time. The interior solution
obtained by an application of the Hern\'andez algorithm, matchs smoothly to
the  M-Q$^{(1)}$ metric on the
boundary of the matter distribution, is regular  and satisfies all standard
physical conditions. In the weak field approximation it represents a
quadrupole
correction to the incompressible fluid sphere.

As we have mentioned before, a spherical core was introduced in
order to assure acceptable physical behaviour at the centre.
 It consists of a sphere of incompressible fluid with possitive
 energy density, larger than pressure
and which matches smoothly to the outer part of  the source.
For the models presented below the spherical core represents
about the $0.1\%$ of the total volume of the source.

Since we are interested in slight deviations from spherical
symmetry we shall consider  models with very small values of $q$. 
Indeed the magnitude of the $q$'s in the models presented are
many orders of magnitude smaller than the values corresponding
to the earth and the sun, which in our units
are approximately $10^{15}$ and $10^5$ respectively \cite{luisyo}.
However for a neutron star  of one solar mass, $10$ km radius
and the same eccentricity as the sun,
the order of magnitude of $q$ is $10^{-4}$ \cite{luisyo},
 as in one of the models below.

It is also worth noticing that the critical value of alfa
($\approx 0.7$) is smaller than the corresponding value in
 the exactly spherically symmetric case (8/9)

We have ran a large number of models for a wide range of values of the
parameters and, both, positive and negative $q$'s. Below we show figures
corresponding to two models, one with
small
$\alpha$ and the other with a value of $\alpha$ close to its critical
value. In both models we considered negative values of $q$ since we are
primarily interested  in oblate objects.

Figures 2--5 exhibit the behaviour of physical variables for an small
value of $\alpha$ (weak field limit). Despite the small value of the
quadrupole parameter, its contribution is
clearly shown in figures 2--4. In figure 5 the numerical error is
comparable with the quadrupole contribution, and accordingly the latter is
somehow screened by the former.

Figures 6-8 display the physical variables for a large value of $\alpha$.
In this case, the quadrupole correction appears sharply in $T^0_{0}$ and
$T^1_{2}$. However it is neglectable
in $T^1_1$, $T^2_2$ and $T^3_3$.

Finally, figure 9 display the range of parameters $\alpha$ and $q$ for
which the solution is physically acceptable.

It is our hope that this source as well as  other interiors to some Weyl
metrics presented elsewhere (\cite{7} and \cite{magli}) could be used as
initial configurations to describe the
departure from equilibrium of very compact objects endowed with an small
but non--vanishing quadrupole structure.

\begin{figure}
\centerline{\epsfxsize=5.in\epsfbox{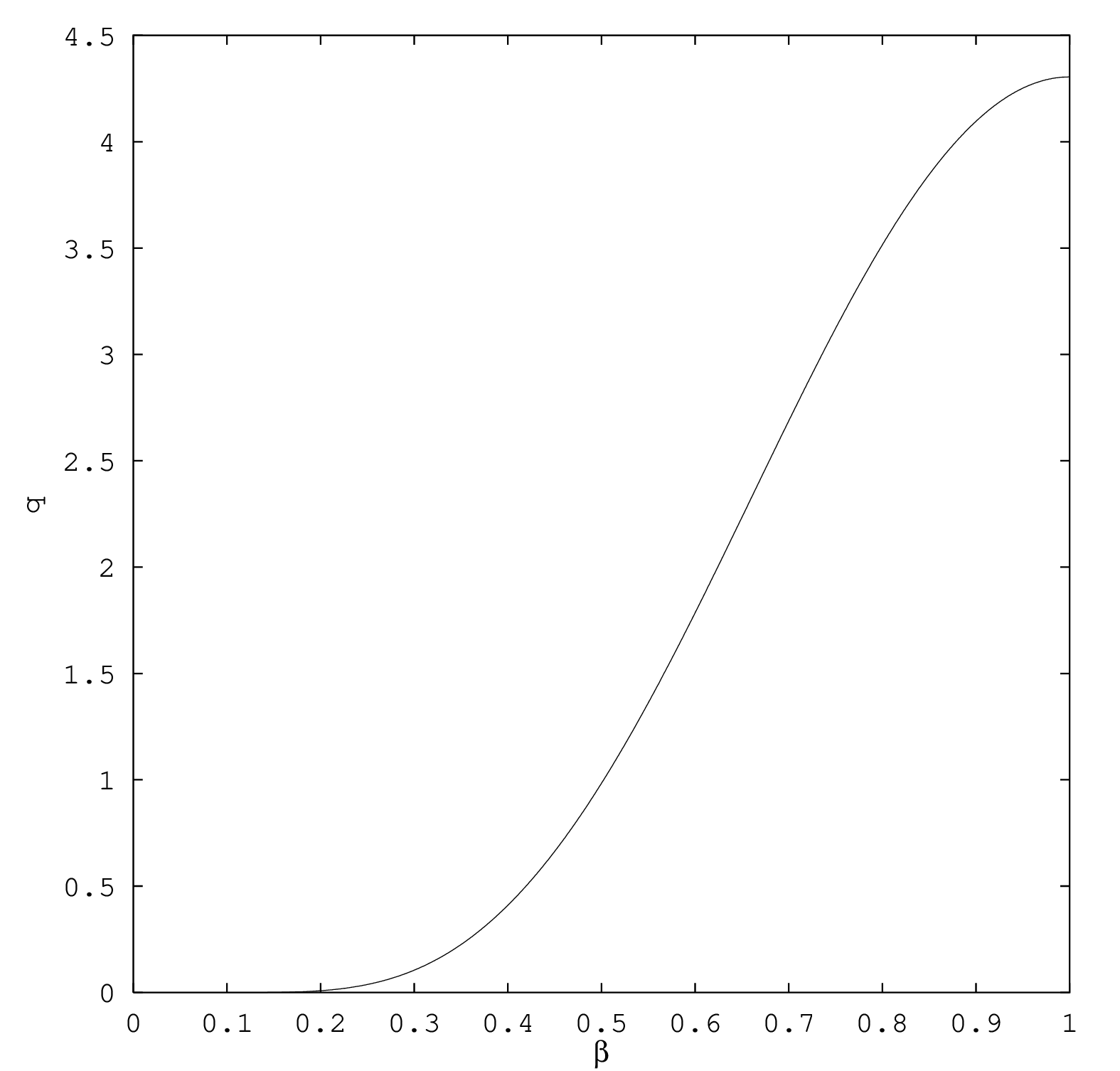}}
\caption{$q$ (multiplied by $10^4$) as a function of $\beta$,
for $\beta_o=0.1$, $\beta_s=1.9$ and $q_o=10^{-3}$.
The maximum value
of $q_{\Sigma}$ is $4.3046721\times 10^{-4}$ which we keep
for the exterior region.}
\label{fig:q}
\end{figure}
\begin{figure}
\centerline{\epsfxsize=4.5in\epsfbox{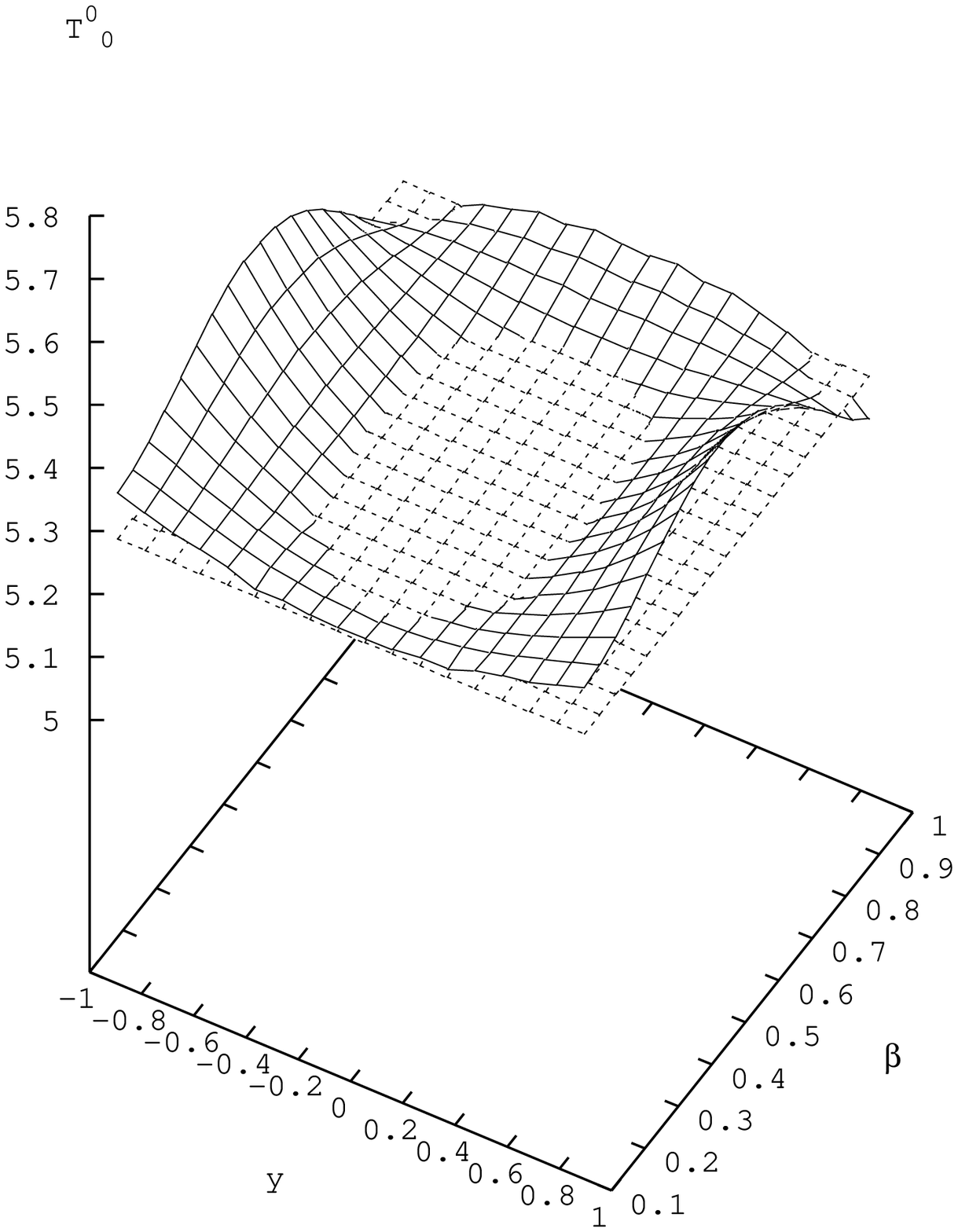}}
\caption{$T^0_0$ (multiplied by $10^7$) as a function of $\beta$ and $y$,
for $\alpha=4/9 \times 10^{-5}$. Dashed lines indicate $q_o=0$
 and continuous lines $q_o=-10^{-12}$.}
\end{figure}

\begin{figure}
\centerline{\epsfxsize=4.5in\epsfbox{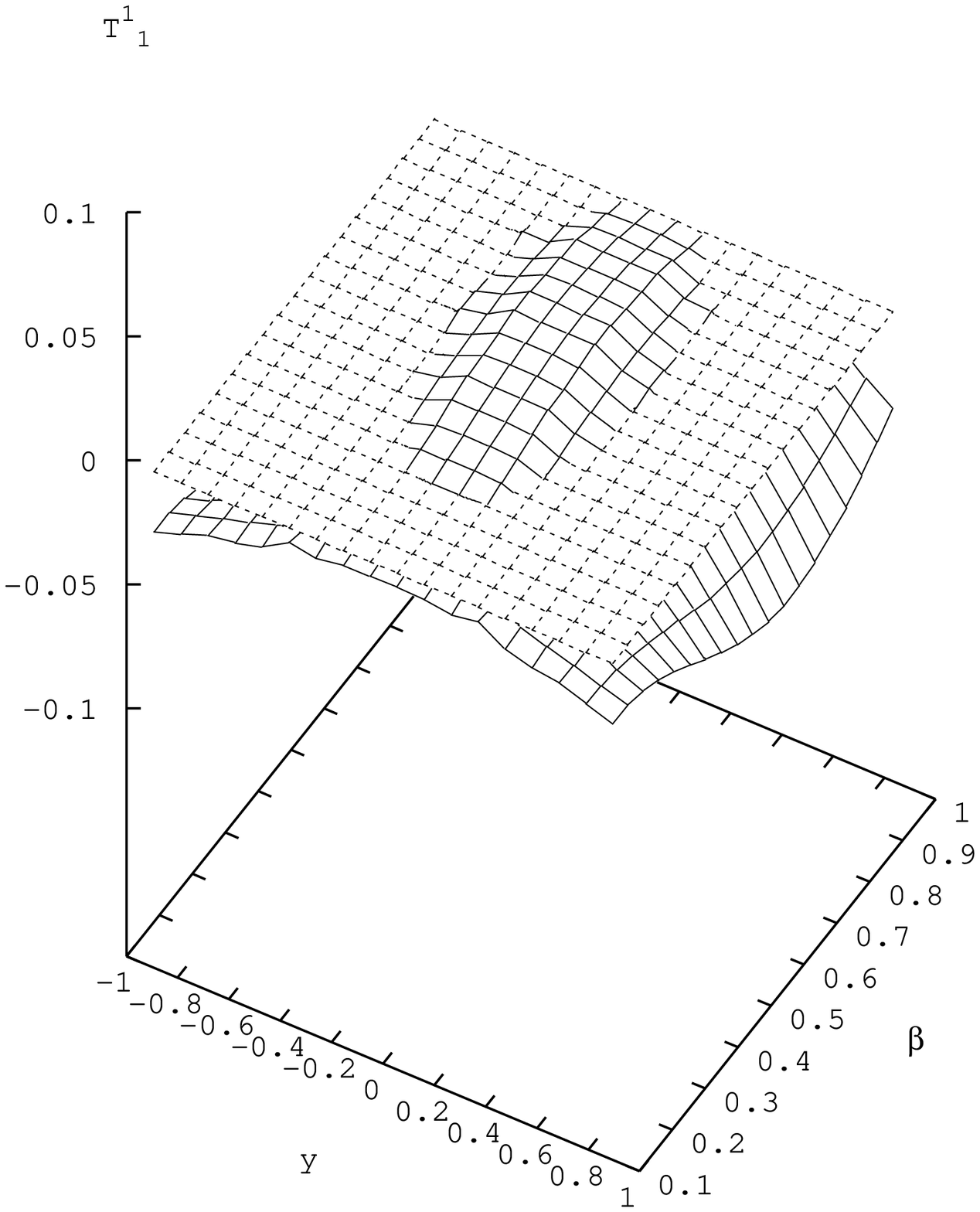}}
\caption{$T^1_1$ (multiplied by $10^7$) as a function of $\beta$ and $y$,
for $\alpha=4/9 \times 10^{-5}$. Dashed lines indicate $q_o=0$
 and continuous lines $q_o=-10^{-12}$.}
\end{figure}

\begin{figure}
\centerline{\epsfxsize=4.5in\epsfbox{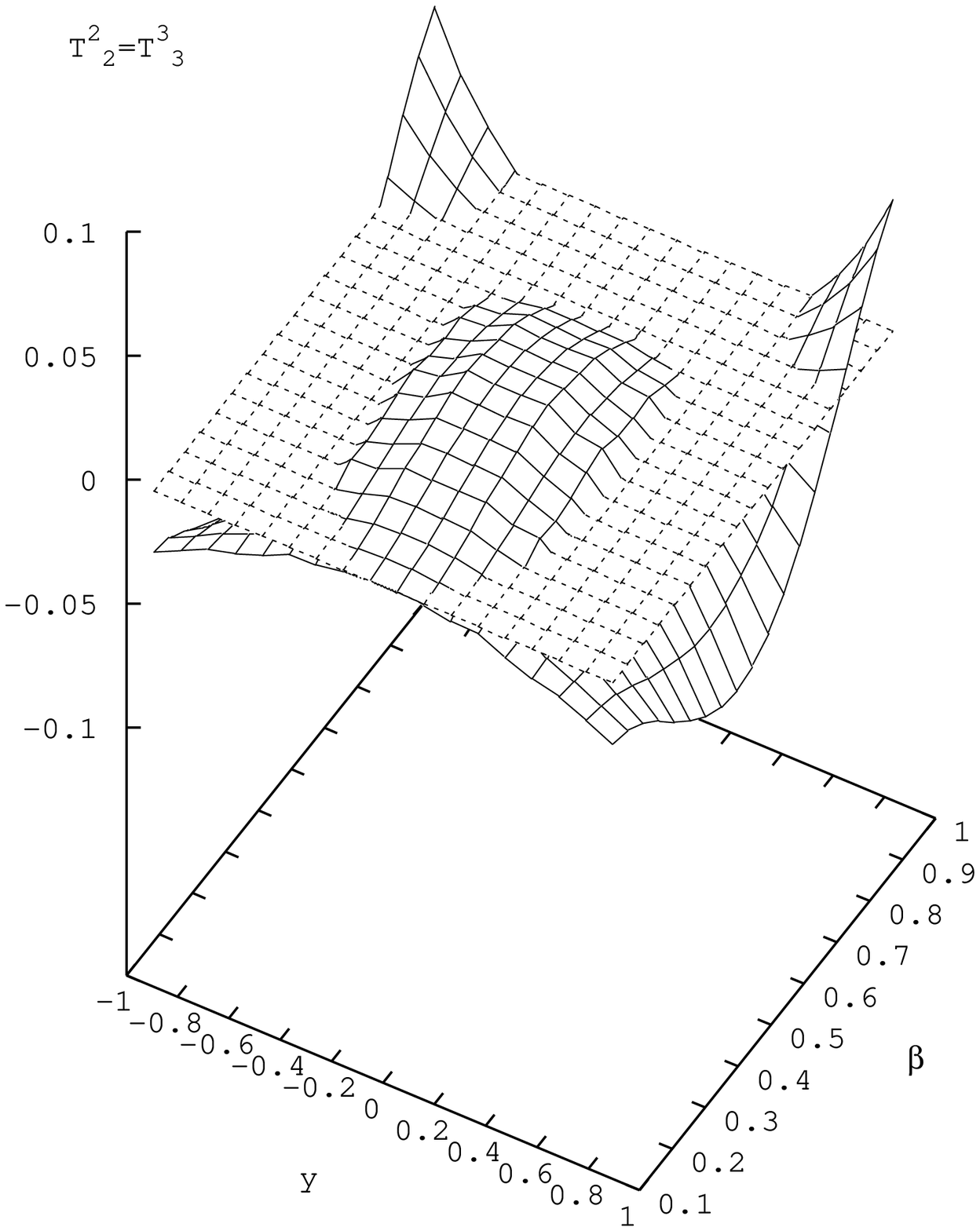}}
\caption{$T^2_2 = T^3_3$ (multiplied by $10^7$) as a function of $\beta$
and $y$,
for $\alpha=4/9 \times 10^{-5}$. Dashed lines indicate $q_o=0$
and continuous lines $q_o=-10^{-12}$.}
\end{figure}
\begin{figure}
\centerline{\epsfxsize=4.5in\epsfbox{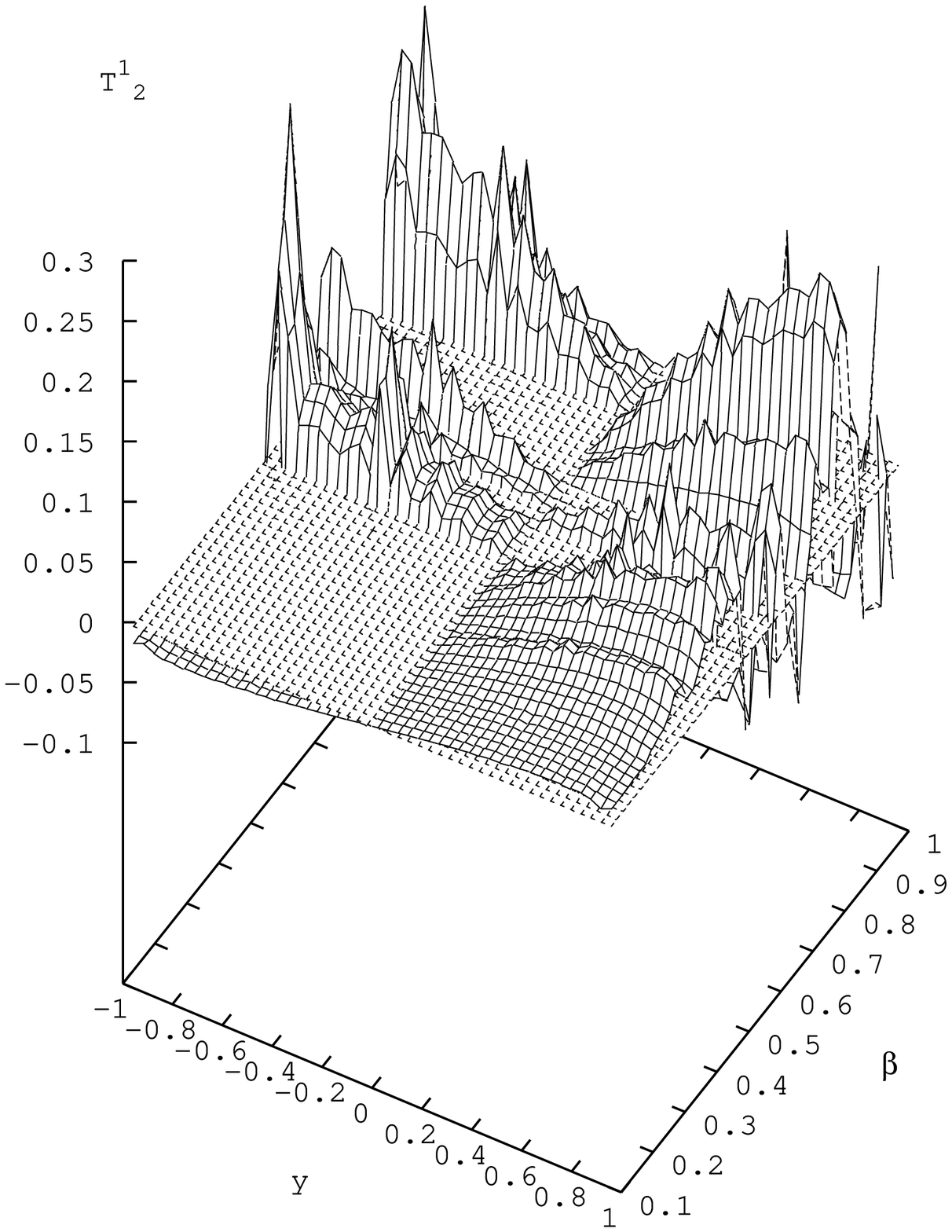}}
\caption{$T^1_2$ (multiplied by $10^{19}$) as a function of $\beta$ and $y$,
for $\alpha=4/9 \times 10^{-5}$. Dashed lines indicate $q_o=0$
and continuous lines $q_o=-10^{-12}$.}
\end{figure}
\begin{figure}
\centerline{\epsfxsize=4.5in\epsfbox{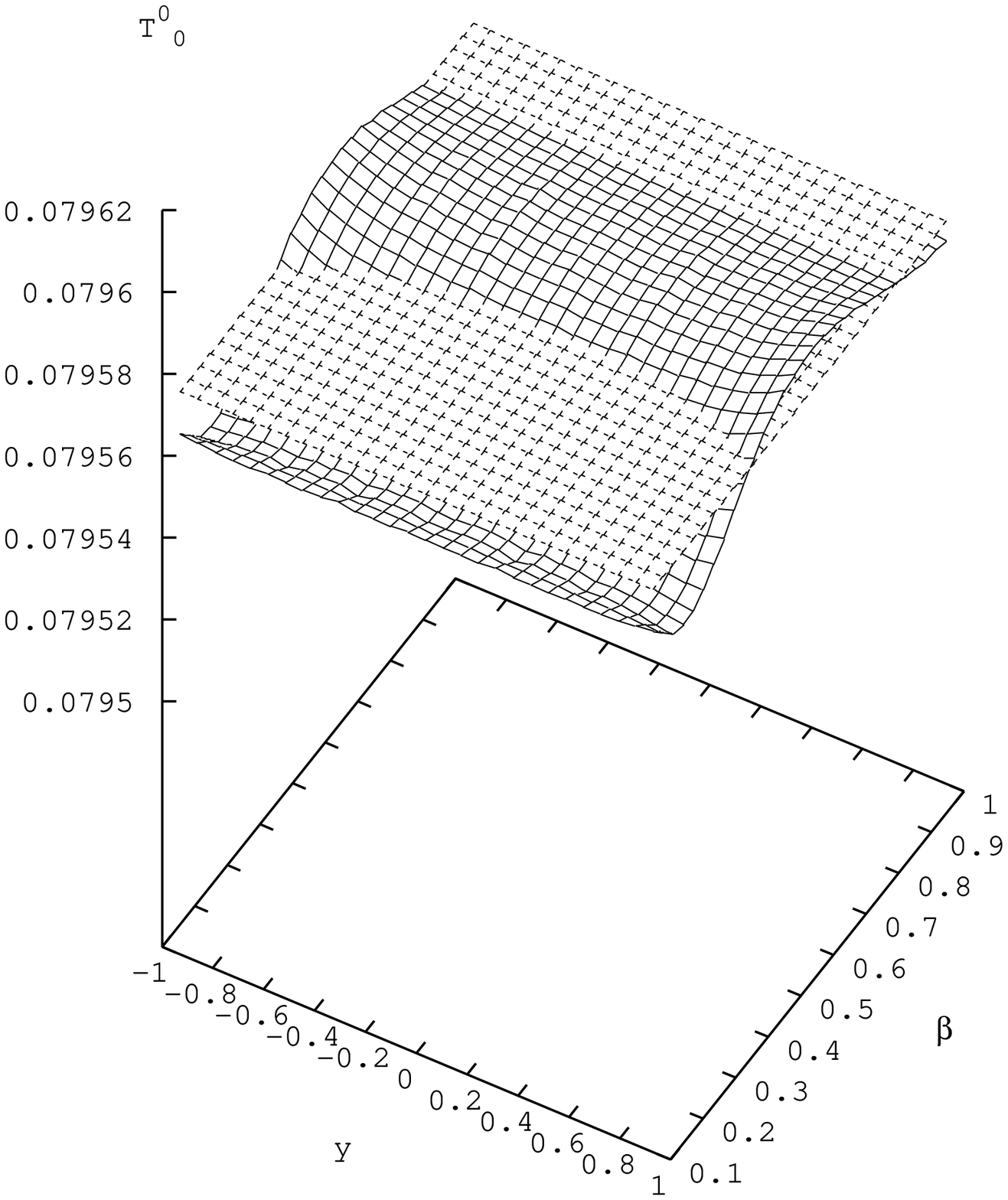}}
\caption{$T^0_0$ as a function of $\beta$ and $y$,
for $\alpha=6/9$. Dashed lines indicate $q_o=0$ and continuous lines
$q_o=-10^{-4}$.}
\end{figure}
\begin{figure}
\centerline{\epsfxsize=4.5in\epsfbox{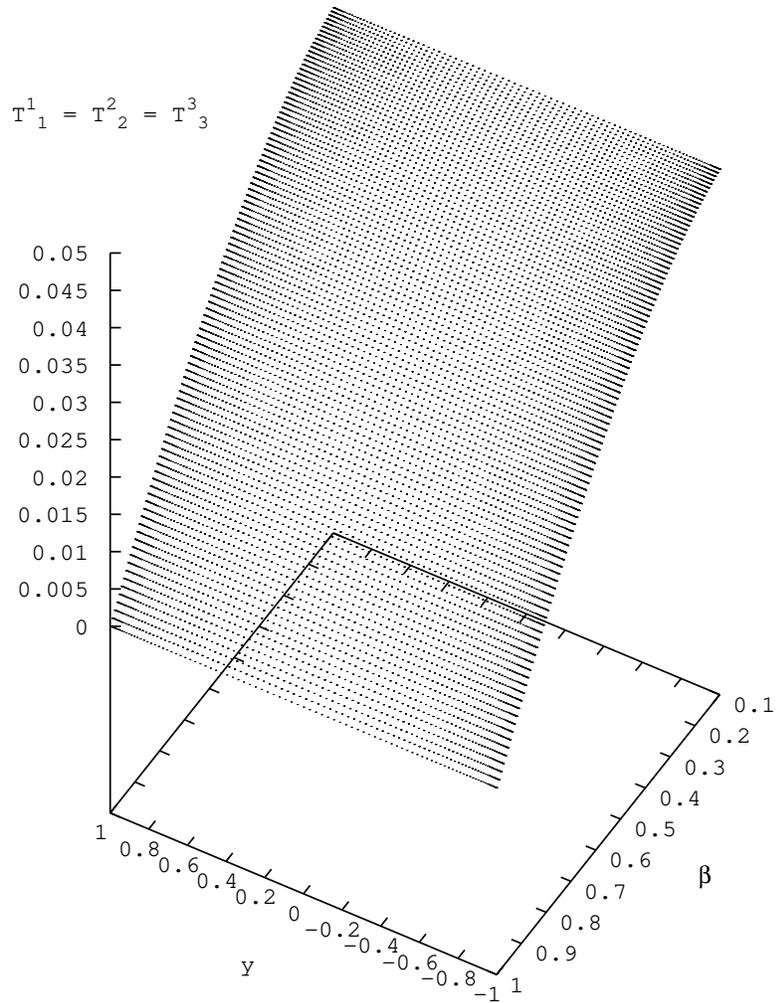}}
\caption{$T^1_1 = T^2_2 = T^3_3$ as a function of $\beta$ and $y$,
for $\alpha=6/9$. All the surfaces overlap for $q_o=0$ and
$q_o=-10^{-4}$.}
\end{figure}
\begin{figure}
\centerline{\epsfxsize=4.5in\epsfbox{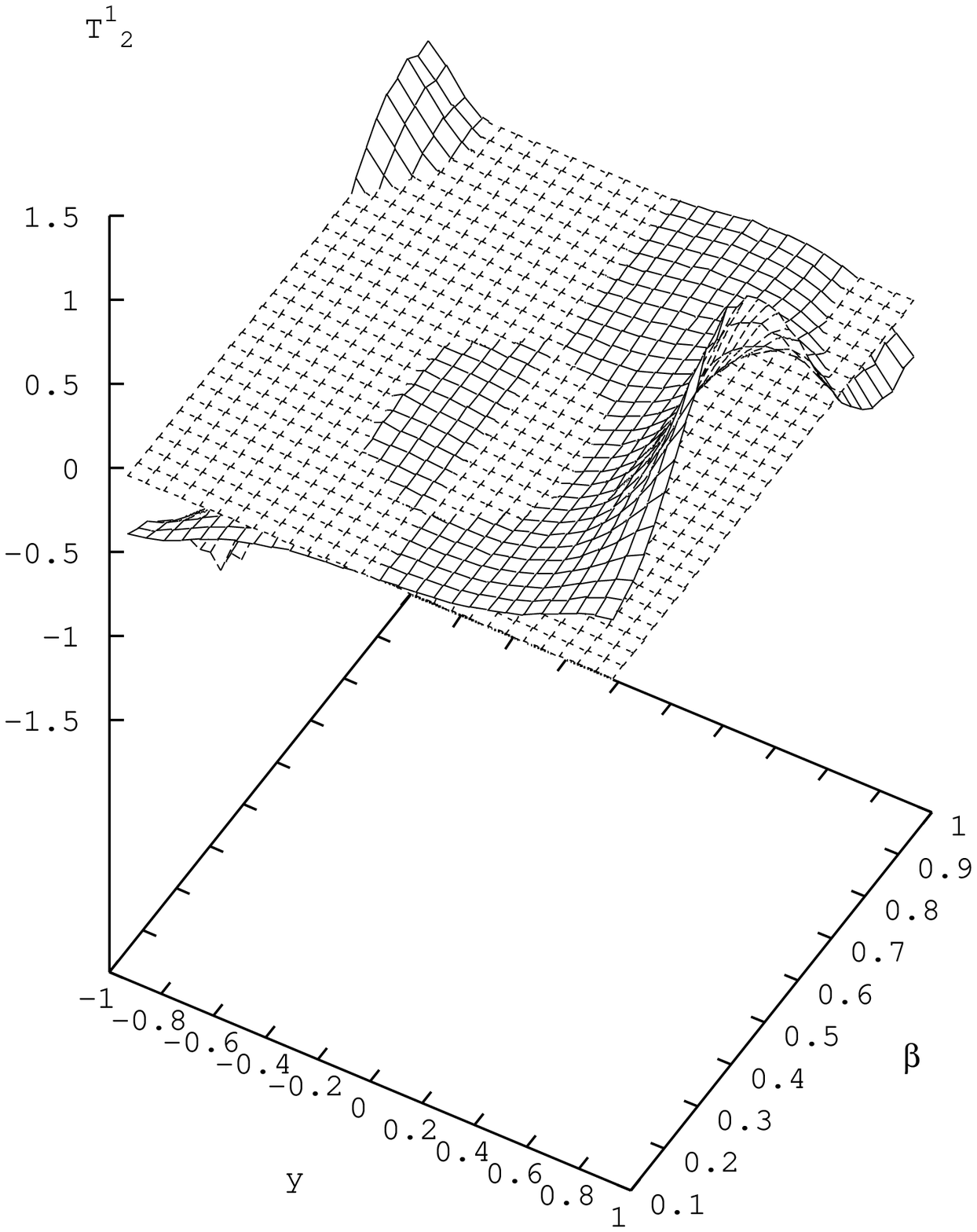}}
\caption{$T^1_2$ (multiplied by $10^{6}$) as a function of $\beta$ and $y$,
for $\alpha=6/9$. Dashed lines indicate $q_o=0$
and continuous lines $q_o=-10^{-4}$.}
\end{figure}
\begin{figure}
\centerline{\epsfxsize=4.5in\epsfbox{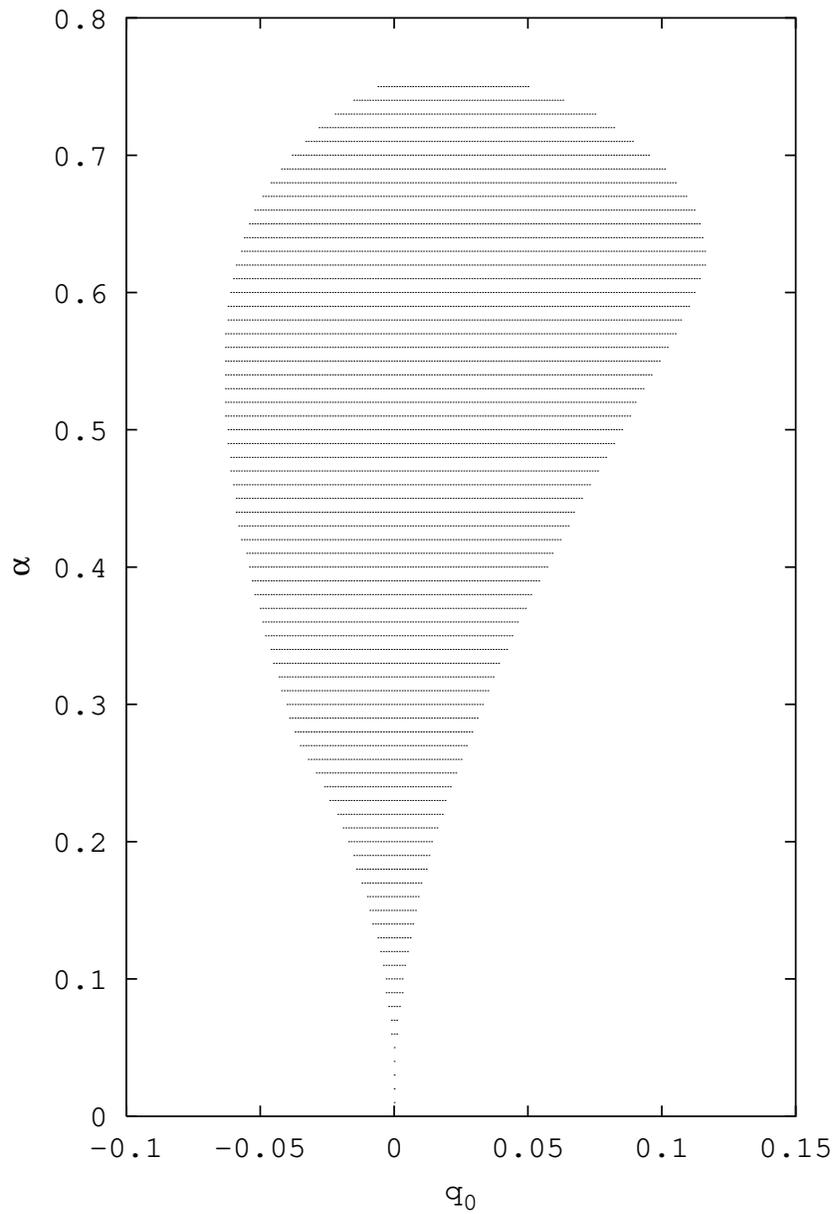}}
\caption{Dashed region
indicates the range of
$\alpha$ and $q_o$ for which models satisfy $T^0_0>0$
and $T^0_0>T^1_1$.}
\end{figure}
\end{document}